\documentclass[pra,twocolumn,showpacs]{revtex4-1}
\usepackage{amsmath,amscd,amsfonts,amssymb,color}
\usepackage{graphicx,amsfonts}
\usepackage[normalem]{ulem}
\usepackage{float}
\usepackage{hyperref}
\newcommand{\ie}{\textit{i}.\textit{e}.}

\begin{document}
\title{Experimental demonstration of selective 
quantum process tomography 
on an NMR quantum information processor} 
\author{Akshay Gaikwad}
\email{akshayg@iisermohali.ac.in}
\affiliation{Department of Physical Sciences, Indian
Institute of Science Education \& 
Research Mohali, Sector 81 SAS Nagar, 
Manauli PO 140306 Punjab India.}
\author{Diksha Rehal}
\email{diksharehal@ug.iisc.in}
\affiliation{Department of Physical Sciences, Indian
Institute of Science Education \& 
Research Mohali, Sector 81 SAS Nagar, 
Manauli PO 140306 Punjab India.}
\author{Amandeep Singh}
\email{amandeepsingh@iisermohali.ac.in}
\affiliation{Department of Physical Sciences, Indian
Institute of Science Education \& 
Research Mohali, Sector 81 SAS Nagar, 
Manauli PO 140306 Punjab India.}
\author{Arvind}
\email{arvind@iisermohali.ac.in}
\affiliation{Department of Physical Sciences, Indian
Institute of Science Education \& 
Research Mohali, Sector 81 SAS Nagar, 
Manauli PO 140306 Punjab India.}
\author{Kavita Dorai}
\email{kavita@iisermohali.ac.in}
\affiliation{Department of Physical Sciences, Indian
Institute of Science Education \& 
Research Mohali, Sector 81 SAS Nagar, 
Manauli PO 140306 Punjab India.}
\begin{abstract}
We present the first NMR implementation of a scheme for selective and
efficient quantum process tomography without
ancilla.  We generalize this scheme such that it can be
implemented efficiently using only a set of 
measurements involving product operators.  The method
allows us to estimate any element of the quantum process
matrix to a desired precision, provided a set of quantum
states can be prepared efficiently.  Our modified
technique requires fewer experimental resources as compared
to the standard implementation of selective and efficient
quantum process tomography,  as it exploits the
special nature of NMR measurements to allow us to compute
specific elements of the process matrix by a restrictive set
of sub-system measurements.
To demonstrate
the efficacy of our scheme, we experimentally tomograph the
processes corresponding to `no operation', a controlled-NOT
(CNOT), and a controlled-Hadamard gate on a two-qubit NMR
quantum information processor, with high fidelities.
\end{abstract} 
\pacs{03.65.Wj, 03.67.Lx, 03.67.Pp} 
\maketitle
\section{Introduction} 
In the last few decades immense progress has been made in the field of quantum
information processing~\citep{nielsen-book-10}.  Although quantum information
processors can be exponentially faster than their classical counterparts, the
working of a quantum computer requires the precise characterization of quantum
states and the ability to perform well-defined quantum operations on them.
While quantum states can be tomographed via state tomography, the actual
physical quantum process that a state undergoes needs to be independently
characterized. Errors may have occurred due to various factors, including
imperfections in the implementation and decoherence processes, leading to a
difference in the actual process as compared to the desired
process~\citep{chuang-jmo-09,poyatos-pra-1997}.  Therefore, it is extremely
important to have experimental protocols which characterize quantum
processes.

Quantum process tomography (QPT)
is a way to characterize
general quantum 
evolutions~\citep{paz-prl-11}. 
The mathematical framework of such a
characterization is based on the fact that any physically valid quantum
dynamics is a completely positive (CP) map and can be expressed as an operator
sum representation.  If we choose a particular operator basis set the map can
in fact be represented  via a process matrix $\chi$.  Hence, the task of the
characterization of a quantum process is equivalent to the $\chi$ matrix
estimation.  This is the standard protocol for QPT.  In order to get a valid
quantum map, the estimated  $\chi$ matrix should be a unit trace, positive
Hermitian operator. For the case when the map does not satisfy these
properties, the maximum likelihood estimation technique can be used to find a
physically valid matrix from the experimental data~\citep{harpreet-pla}.
While the maximum likelihood estimation method always yields a
valid density matrix, care should be taken when estimating special states
such as entangled states,
as recent discussions
have showed that the method can lead to systematic errors
in such cases~\cite{mle-prl,mle-bias}.

QPT has been extensively used in characterizing quantum
decoherence~\citep{poyatos-opt-1998,hegde-pra-14,emerson-sci-07,kofman-pra-09}
and various quantum
gates~\citep{mahesh-pra-14,obrien-prl-04,weinstein-jcp-04}.
Its potential application has been exploited in developing
quantum error correction codes~\citep{lopez-pra-09} and
estimating Lindblad operators and master equation parameters
for a noisy
channel~\citep{howard-njp-06,ofek-nature-16,gaussian}.  The
physical realization of QPT has been demonstrated on
different experimental setups such as
NMR~\citep{childs-pra-01,maciel-njp-15}, superconducting
qubits~\citep{neeley-nature-08,chow-prl-09,bialczak-np-10,dewes-prl-12},
nitrogen vacancy centers~\citep{zhang-prl-14}, linear
optics~\citep{martini-pra-03} and ion-trap based quantum
processors~\citep{riebe-prl-06}.

The complete characterization of the quantum process based
on the standard QPT protocol is experimentally as well as
computationally a daunting task, as it requires  high-cost state
tomographs~\citep{leskowitz-pra-04,lidar-pra-08}.  Several
attempts have been made in the past few years to simplify
the QPT protocol, which involve prior knowledge about the
commutation relations of the system Hamiltonian and
the system-environment interaction Hamiltonian~\citep{z-jcp-13},
performing ancilla-assisted
tomography~\citep{altepeter-prl-03}, using
techniques of direct characterization of quantum
dynamics~\citep{mohseni-prl-06,mohseni-pra-07} and process
tomography via adaptive measurements~\citep{jefundu-sci-16}.
Although these methods offer some advantages over standard
QPT, they still are not very useful when only certain
elements of the $\chi$ matrix need to be estimated.  Hence
much effort has recently focused on achieving a selective
estimation of elements of the $\chi$ matrix via a technique
called selective and efficient quantum process tomography
(SEQPT) without
ancilla~\citep{paz-prl-08,paz-pra-09,paz-prl-10}.
The SEQPT without ancilla method interprets the elements of
the $\chi$ matrix as an average of the survival
probabilities of a certain quantum map; while the method
certainly has advantages over other existing schemes, it
still requires a large number of state preparations and
experimental settings to carry out complete process
tomography.

In this work, we propose a generalization of
the SEQPT method without ancilla, which requires fewer
experimental resources as compared to the SEQPT or the
standard QPT protocols.  We exploit the fact that the density
operator proportional to identity does not produce any NMR
signal and use the product operator formalism to achieve
selective estimation of the quantum process matrix to a
desired precision.  We call our scheme modified selective
and efficient quantum process tomography (MSEQPT). 
Our scheme achieves a simplification of the QPT protocol as
in this  scheme, the
detection settings need not be changed each time while
estimating different elements of $\chi$ matrix. Our scheme
is efficient as it relies on calculating the expectation
values of special Hermitian observables by 
locally measuring the expectation values of the basis
operators in a pre-decided set of quantum states. We
experimentally demonstrate our scheme by implementing it on
a two-qubit NMR system, where we tomograph the `no
operation', the controlled-NOT and the controlled-Hadamard
gates.

This paper is organized as follows:~In Section~\ref{MSEQPT}
we detail the MSEQPT scheme where we use product operators
and NMR measurements to implement SEQPT.
Section~\ref{2-design} gives details of the quantum 2-design
used in the experiments.  Section~\ref{NMRImp} contains a
description of the experimental implementation of the
MSEQPT scheme on a two-qubit NMR quantum information
processor.  The results of the quantum processes that were
experimentally tomographed are presented in the later part
of Section~\ref{NMRImp}.  Section~\ref{Con} contains a few
concluding remarks.
\section{Selective and efficient quantum process tomography 
using product operators measurements}
\label{MSEQPT}
Consider a quantum system undergoing a general quantum
evolution represented by a completely positive (CP) map.
The action of such a  map on a quantum state $\rho$ via the
superoperator $\Lambda$ in the operator sum representation
is given as follows:
\begin{equation}
\Lambda(\rho)=\sum_{i} A_i \rho A_i^\dagger ;\quad \quad \sum_{i}
A_i A_i^\dagger = I.
\label{cpmap}
\end{equation}
where $A_{i}$'s are the Kraus
operators~\citep{sudarshan-pr-61,kraus-book-83}.   Consider a
complete set of $D^2-1$ basis operators $\lbrace E_i\rbrace$
for a $D$-dimensional Hilbert space, 
satisfying the 
orthogonality conditions
\begin{equation}
\textbf{Tr}(E_m E_n^\dagger)=D\delta_{mn}\,\,{\rm and}\,\,
\textbf{Tr}(E_m)=D\delta_{m0}.
\label{ortho}
\end{equation}
In term of these operators the
Kraus operators can be expanded as $A_i=\sum_m a_{im}E_m$
and 
any CP map given in Eq.~(\ref{cpmap})  can be rewritten
as follows~\citep{chuang-jmo-09}:
\begin{equation} 
\label{e_1.1} 
\Lambda(\rho)=\sum_{a,b} \chi_{ab} {E_a} \rho
{E_b}^\dagger 
\end{equation}
where the quantities $\chi_{ab}$ are the elements of a
matrix $\chi$ characterizing the given CP map $\Lambda$. This
is known as the $\chi$ matrix representation of the quantum
process.

A major step towards the determination of $\chi_{ab}$
elements is to relate them to the quantities $F_{ab}$ called
the average survival probabilities~\citep{paz-pra-09,paz-prl-11}:
\begin{equation}
\label{e_3} F_{ab} \equiv \frac{1}{K}
\sum_{j} \langle \phi_j \vert \Lambda (E_a^\dagger \vert
\phi_j \rangle \langle \phi_j \vert E_b)\vert \phi_j \rangle
=\frac{ D\chi_{ab}+\delta_{ab}}{D+1} 
\end{equation} 
Here a quantum 2-design set $S=\{ \vert \phi_j\rangle:
j=1,....,K \}$ of cardinality $K$ has been used to  provide
a way to discretely sample the system Hilbert space so as to
avoid integration over the entire
space~\citep{dankert-pra-09}.  Thus by
evaluating the summation given in Eq.~(\ref{e_3}) for a
given $a$ and $b$, one can selectively find the matrix
element $\chi_{ab}$.

The operator 
$E^{\dagger}_{a}\vert \phi_j\rangle\langle \phi_j \vert E_{b}
=E^{\dagger}_{a} \rho_j E_{b}$, in general is not
a valid density operator (unless $E_a=E_b$) and hence cannot
be created in an experiment, thus preventing the
determination of $F_{ab}$. Extensions involving valid
density operators of the form
$\Lambda ((E_a \pm E_b)^\dagger \rho_j 
(E_a \pm E_b))$ have been proposed to circumvent this
problem and determine the probabilities $F_{ab}$
experimentally~\citep{paz-prl-11}.  However, these
procedures involve using different experimental settings for
different values of $a$'s and $b$'s to prepare the required
state and a large number of experiments have to be performed in
order to achieve a high precision. Further, constructing and
implementing the corresponding unitary operators is a
challenging task.

We take a different approach to implement SEQPT using
a method where we take weighted average  results of
different experiments analogous to
the temporal averaging scheme to obtain a pseudopure 
state~\citep{vandersypen-prl-00,leung-rsl-98}.
In this way we compute the expectation values of basis
operators by an appropriate mapping of the desired measurements 
onto measurements of individual spin magnetizations.
Eq.~(\ref{e_3}) can be rewritten in terms of density
operators corresponding to the quantum 2-design states $\rho_j=\vert
\phi_j \rangle \langle \phi_j\vert$ as:
\begin{equation}
\label{e_2.5}
F_{ab} \equiv \frac{1}{K} \sum_{j} \textbf{Tr} [ \rho_j
\Lambda (E_a^\dagger \rho_j E_b)]=\frac{
D\chi_{ab}+\delta_{ab}}{D+1}
\end{equation}
The basis operators  $\lbrace E_i \rbrace$ can be used to 
decompose the operator $E^{\dagger}_a \rho_j E_b$:
\begin{equation}
\label{e_2.6} 
E_a^\dagger \rho_j E_b=
\sum_{i} {^jc}_i^{ab}E_i 
\end{equation}  
where the coefficients $^jc_i^{ab}\in\mathbb{C}$ are
independent of the quantum process characterized by $\Lambda$,
and can be computed analytically using the orthogonality
condition: 
\begin{equation}
{^jc_i}^{ab}=
\frac{1}{D}\textbf{Tr}[(E_a^\dagger \rho_j E_b) E_i]
\label{coefficients}
\end{equation}
The superoperator $\Lambda$ is linear and hence can be
expanded as:
\begin{equation}
\label{e_2.7}
\Lambda (E_a^\dagger \rho_j E_b)=  \sum_{i} {^jc}_i^{ab} \Lambda(E_i)
\end{equation}
Using the above decomposition, Eq.(\ref{e_2.5}) can be rewritten as
\begin{equation}\label{e_2.8}
F_{ab}= \frac{1}{K} \sum_{j} \textbf{Tr}\left[\rho_j \sum_{i}
{^jc}_i^{ab} \Lambda(E_i)\right] 
\end{equation} 
Every  
basis operator $E_i$  (other than the first one which we
take proportional to identity) is a  
Hermitian operator with zero trace; we can interpret it as a
deviation density operator and can make it unit trace by
adding identity divided by the dimension, and thus it can be
experimentally prepared as a valid  quantum state.
For our purpose since we work
with NMR quantum information processors, the addition of multiples of identity
does not contribute to the NMR signal and therefore such terms can be
ignored.
The quantum process $\Lambda$ can
then be allowed to act on this basis operator state giving
us $\Lambda (E_i)$ for every basis vector. Therefore if
we tomograph the state $\Lambda (E_i)$ experimentally, we
can use the theoretically calculated coefficients
${^jc_i}^{ab}$ as per the Eq.~(\ref{coefficients}) and compute
$F_{ab}$ in Eq.~(\ref{e_2.8}). 
The
results from individual $E_i$'s weighted by $^jc_i^{ab}$ are added to
obtain the final result.

Our aim is to avoid the full state tomography of the state
$\Lambda(E_i)$.  Decomposing $\rho_j$ as $\rho_j=\sum_{k}
{^je_k E_k}$ (with $^je_k\in\mathbb{R}$), and using the
linearity of trace, Eq.~(\ref{e_2.8}) reduces to 
\begin{equation}\label{e_2.10}
F_{ab}= \frac{1}{K} \sum_{j} \sum_{i,k} {}^je_k\, {}^jc^{ab}_i
\textbf{Tr} [E_k \Lambda(E_i)]
\end{equation}
where the coefficients ${}^je_k \,{}^jc^{ab}_i$ are process 
independent and can
be computed analytically. Rewriting them as $^j\beta^{ab}_{ki}$,
Eq. (\ref{e_2.10}) takes a simple form:
\begin{equation}\label{e_2.11}
F_{ab}= \frac{1}{K} \sum_{i,j,k} {^j\beta^{ab}_{ki}} \textbf{Tr}
\left[E_k \Lambda(E_i)\right]
\end{equation}
where $\textbf{Tr}[E_k \Lambda(E_i)] \equiv \langle E_k^i
\rangle$ is the expectation value of basis operator $E_k$ in
the state $\Lambda(E_i)$.  The information about the quantum
process is now
stored in the output state $\Lambda(E_i)$. To calculate a selective
element of $\chi$ matrix, all we need to do is to calculate
expectation values of $E_k$ and take the weighted average of
these expectations using the theoretically calculated 
coefficients $^j\beta^{ab}_{ki}$.

To determine $F_{ab}$, we need not perform full quantum
state tomography of the output state $\Lambda(E_i)$ which is
a very expensive operation.  The expectation values $\langle
E^i_k \rangle$ can be determined by mapping them to 
expectation values of appropriate single-spin operators.  To demonstrate
this we choose the Pauli basis as our $\lbrace E_i\rbrace$ which
for $N-$qubits involves choosing $\{I^j, \sigma_x^j,
\sigma_y^j, \sigma_z^j \}$ for the $j$th qubit  and
taking all possible tensor products to form the set
$\{E_i\}$. The measurements of elements of the Pauli basis can
be measured via  individual spin measurements and in fact can be 
mapped to measurements of various $\sigma_z^j$ by applying 
certain fixed operations before measurement. This is
particularly suitable for NMR where such measurements can be
readily accomplished.
For a two-qubit system 
this map is given in Table~\ref{Map}
where the measurement of each member of the Pauli basis set is 
mapped to a measurement of certain 
single-spin $z$ magnetizations.
This significantly simplifies the experimental complexity
of the SEQPT scheme.
\begin{figure}[t]
\includegraphics[angle=0,scale=1]{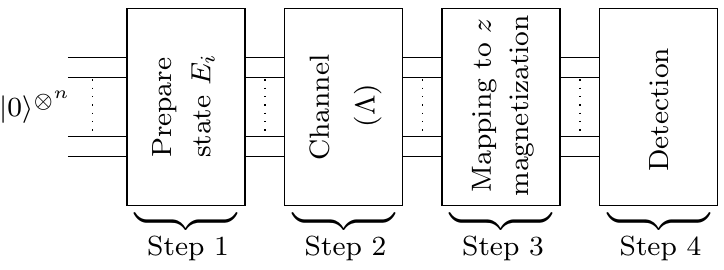}
\caption{Block diagram of MSEQPT protocol.
Step 1: Preparation of basis operator $E_i$ state. Step 2:
Application of quantum channel $\Lambda$.  Step 3: Mapping
of basis operator to  measurements  of individual
spin magnetizations followed
by expectation value measurements of Pauli
$z$-operators on subsystems.  Step 4: 
Detection on individual spins.
}
\label{SEQPTlm}
\end{figure} 
A stepwise description of the experimental implementation of
the MSEQPT protocol to selectively determine the element
$\chi_{ab}$ of the process matrix is as follows:
\begin{table}[h] \caption{\label{Map} Fifteen observables
for two qubits mapped to  
\textit{z}-magnetization of one of the  qubits.  $\rho_k=U_k
\Lambda(E_i)U_k^\dagger$ in order to calculate $\langle
E_k\rangle=\textbf{Tr}\left[E_k\Lambda(E_i)\right]$.}
\begin{ruledtabular}
\begin{tabular}{rp{12pt}l}
\textrm{Observable Expectation}&&
\textrm{Unitary operator $U_k$}\\
\colrule
$\langle \sigma_{2x} \rangle$ = Tr[$\rho_{1}.\sigma_{2z}$]
&&
$U_{1}=\overline{Y}_2$  \\
$\langle \sigma_{2y} \rangle$ = Tr[$\rho_{2}.\sigma_{2z}$]
&&
$U_{2}=X_2$ \\ 
$\langle \sigma_{2z} \rangle$ = Tr[$\rho_{3}.\sigma_{2z}$] &&
$U_{3}$=Identity  \\ 
$\langle \sigma_{1x} \rangle$ =
Tr$\left[\rho_{4}.\sigma_{1z}\right]$ &&
$U_{4}=\overline{Y}_1$ \\ 
$\langle \sigma_{1x}\sigma_{2x} \rangle$ = Tr[$\rho_{5}.
\sigma_{2z}$] && $U_5={\rm{CNOT~  }}Y_2Y_1$  \\ 
$\langle \sigma_{1x}\sigma_{2y} \rangle$ =
Tr[$\rho_{6}.\sigma_{2z}$] && $U_6={\rm{CNOT~  }}\overline{X}_2Y_1$
\\ 
$\langle \sigma_{1x}\sigma_{2z} \rangle$ =
Tr[$\rho_{7}.\sigma_{2z}$] && $U_7={\rm{CNOT~  }}\overline{Y}_1$ \\ 
$\langle \sigma_{1y} \rangle$ = Tr[$\rho_{8}.\sigma_{1z}$]
&&
$U_{8}=X_1$ \\ 
$\langle \sigma_{1y}\sigma_{2x} \rangle$ =
Tr[$\rho_{9}.\sigma_{2z}$] && $U_9={\rm{CNOT~  }}\overline{Y}_2X_1$
\\ 
$\langle \sigma_{1y}\sigma_{2y} \rangle$ =
Tr[$\rho_{10}.\sigma_{2z}$] &&
$U_{10}={\rm{CNOT~  }}\overline{X}_2\overline{X}_1$  \\ 
$\langle \sigma_{1y}\sigma_{2z} \rangle$ =
Tr[$\rho_{11}.\sigma_{2z}$] && $U_{11}={\rm{CNOT~  }}X_1$  \\ 
$\langle \sigma_{1z} \rangle$ = Tr[$\rho_{12}.\sigma_{1z}$]
&& $U_{12}$=Identity \\ 
$\langle \sigma_{1z}\sigma_{2x} \rangle$ = 
Tr[$\rho_{13}.\sigma_{2z}$] && $U_{13}={\rm{CNOT~  }}\overline{Y}_2$  \\ 
$\langle \sigma_{1z}\sigma_{2y} \rangle$ = 
Tr[$\rho_{14}.\sigma_{2z}$] && $U_{14}={\rm{CNOT~  }}X_2$ \\ 
$\langle \sigma_{1z}\sigma_{2z} \rangle$ = 
Tr[$\rho_{15}.\sigma_{2z}$] && $U_{15}=\rm{CNOT~  }$ \\ 
\end{tabular}
\end{ruledtabular}
\end{table}
\leftmargini=16pt
\begin{enumerate}
\item[(i)] Choose any state $\rho_j=\vert \phi_j\rangle
\langle \phi_j\vert$ from the
set of quantum 2-design and find the decomposition  of
$E_a^\dagger \rho_j E_b $ in terms of basis operators $E_i$.
\item[(ii)]
Experimentally prepare the quantum system in one of the  basis
states having non-vanishing coefficients ${}^jc^{ab}_i$ as per
Equation~(\ref{e_2.6}).
\item[(iii)] Apply the quantum channel $\Lambda$ to $E_i$ to get
the output state $\Lambda(E_i)$. 
\item[(iv)] Find the decomposition of the chosen state $\rho_j$
in terms of basis operators analytically and then
experimentally determine the expectation values of all those
$E_k$'s which have
non-vanishing coefficients, ${}^je_k$, using 
our
measurement technique. 
\item[(v)] Repeat the procedure for all the states in the
chosen  quantum
2-design set.
\end{enumerate}
The MSEQPT protocol is schematically depicted in
Fig.~\ref{SEQPTlm}: the first step is to prepare the basis
state, followed by the action of the quantum process. After
the quantum process has acted on the basis state the next
step is to map the required measurements to 
single-spin magnetization measurements and finally we do the 
single spin magnetization detection.

Our modified scheme has two advantages: first, it is
simpler than the original scheme as we do not have to choose
different experimental settings for the estimation of each
element of the $\chi$ matrix and second, it involves fewer
experiments. The  comparison of experimental resources 
required by different protocols to determine a specific
element of the $\chi$ matrix for two-qubit systems is
given in Table~\ref{complexity}.
\begin{table}[h!]
\caption{\label{complexity}
Comparison of experimental resources for different
protocols for the determination of a specific element of
the $\chi$ matrix for a two-qubit system}
\begin{ruledtabular}
\begin{tabular}{|c|c|c|c|}
\hline
&QPT & SEQPT & MSEQPT \\
\hline
Preparations &15&80&15\\
\hline
Readouts & 120&240& 60\\
\hline
\end{tabular}
\end{ruledtabular}
\end{table}
The standard QPT method implemented on two NMR qubits relies on the
channel action given by  
\begin{equation}
 \rho_{out}=\sum_{a,b} \chi_{ab} {E_{a}}\rho_{in} {E_{b}^{\dagger}}
\end{equation}
This requires 
state preparation settings ($\rho_{in}$)= 15, with the
number of tomographs required being 15. Since each tomograph
requires 8 readouts, the total number of readouts required
is $ 15 \times 8 =120$. 

In the SEQPT protocol, the states to be prepared for
estimating the real part of $\chi_{ab}$ are: $(E_a \pm
E_b)^\dagger \rho_j ((E_a \pm E_b))$, where j=1 to 20
(2-design states).  The states to be prepared for estimating
the imaginary part of $\chi_{ab}$ are: $(E_a \pm
iE_b)^\dagger \rho_j ((E_a \pm iE_b))$.  The number of state
preparation settings required to obtain $\chi_{ab}$ = 80 (20
for $(E_a + E_b)$ + 20 for $(E_a - E_b)$ + 20 for $(E_a +i
E_b)$ + 20 for $(E_a - iE_b)$).  This method requires 3
readouts (the number of non vanishing coefficients in the
expansion of $\rho_j$) for each state, in order  to obtain
transition probabilities $\textbf{Tr} [ \rho_j \Lambda ((E_a
\pm E_b)^\dagger \rho_j (E_a \pm E_b))]$ and $\textbf{Tr} [
\rho_j \Lambda ((E_a \pm iE_b)^\dagger \rho_j (E_a \pm
iE_b))]$.  Hence the total number of readouts required
required for the SEQPT method is $3\times 80 = 240$.  For
the MSEQPT protocol, the number of state preparation
settings are 15 while the number of readouts required for
each state preparation is 4.  Hence the total number of
readouts required required for the MSEQPT method is $4
\times 15 = 60$.
\subsection{Quantum 2-design set using mutually
unbiased basis} \label{2-design}
One of the requirements for experimental implementation of
MSEQPT is the quantum 2-design set $S$, and algorithms
to construct such a set are
available~\citep{dankert-pra-09,paz-pra-09}.  One way is to
find a complete set of mutually unbiased basis (MUBs) states
where a  system with a $D$  dimensional state space will
have ($ D+1 $) MUBs if $D$ is a prime number or power of a prime
number~\citep{bandyopadhyay-alg-02,lawrence-pra-02,
Klappenecker-book-04}.  For our two-qubit system $D=2^2$, and the
set of quantum 2-design can be constructed by using a
complete set of MUBs, which are five in this case.  The MUBs
states satisfy the relation, ${\vert \langle \phi_p^{B_k}
\vert \phi_q^{B_l} \rangle\vert}^2=\frac{1}{D}$ for all $k
\neq l$, $B_k$'s are basis set labels and $\phi_p$'s are the
elements within the basis set. The complete set of MUBs
constituting states in quantum 2-design set for 2-qubit
system $(D=4)$, in the computational basis, is given below
\citep{Klappenecker-book-04}:
\begin{equation*}
B_1=\left \lbrace \begin{pmatrix}
1\\0\\0\\0
\end{pmatrix},\begin{pmatrix}
0\\1\\0\\0
\end{pmatrix},\begin{pmatrix}
0\\0\\1\\0
\end{pmatrix},\begin{pmatrix}
0\\0\\0\\1
\end{pmatrix}\right \rbrace 
\end{equation*}
\begin{equation*}
B_2=\frac{1}{2}\left \lbrace \begin{pmatrix}
1\\1\\1\\1
\end{pmatrix},\begin{pmatrix}
1\\1\\-1\\-1
\end{pmatrix},\begin{pmatrix}
1\\-1\\-1\\1
\end{pmatrix},\begin{pmatrix}
1\\-1\\1\\-1
\end{pmatrix}\right \rbrace 
\end{equation*}
\begin{equation*}
B_3=\frac{1}{2} \left \lbrace \begin{pmatrix}
1\\i\\i\\-1
\end{pmatrix},\begin{pmatrix}
1\\-i\\i\\1
\end{pmatrix},\begin{pmatrix}
1\\i\\-i\\1
\end{pmatrix},\begin{pmatrix}
1\\-i\\-i\\-1
\end{pmatrix}\right \rbrace  
\end{equation*}
\begin{equation*}
B_4=\frac{1}{2} \left \lbrace\begin{pmatrix}
1\\-1\\-i\\-i
\end{pmatrix},\begin{pmatrix}
1\\-1\\i\\i
\end{pmatrix},\begin{pmatrix}
1\\1\\-i\\-i
\end{pmatrix},\begin{pmatrix}
1\\1\\-i\\i
\end{pmatrix}\right \rbrace 
\end{equation*}
\begin{equation}
B_5=\frac{1}{2}\left \lbrace \begin{pmatrix}
1\\-i\\-1\\-i
\end{pmatrix},\begin{pmatrix}
1\\-i\\1\\i
\end{pmatrix},\begin{pmatrix}
1\\i\\-1\\i
\end{pmatrix},\begin{pmatrix}
1\\i\\1\\-i
\end{pmatrix}\right \rbrace
\end{equation}
For example $\vert \phi_3^{B_1} \rangle $ is the third
element of $B_1$ basis set and the state is $\vert 10 \rangle$.
Also $B_1$ is the commonly used computational basis. All the
twenty states in the above defined MUBs comprise the quantum
2-design set $S$ for the present study.
\begin{figure}[h]
\includegraphics[angle=0,scale=1]{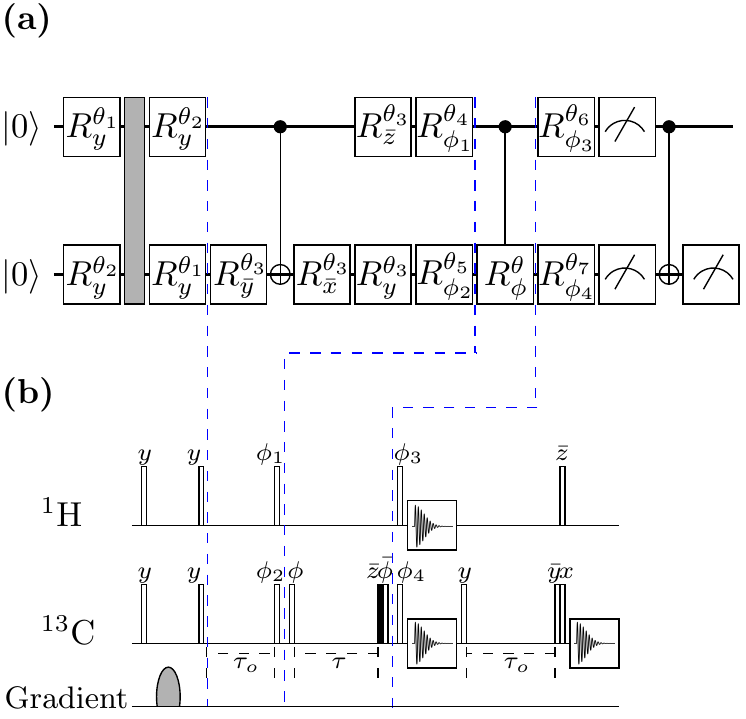}
\caption{(a) Quantum circuit to implement MSEQPT. 
Local unitary operations $R_{\phi}^\theta$  
are achieved via rotations by an angle $\theta$ and
phase $\phi$. The values of the angles $\theta_l$ are either $0$ or 
$\frac{\pi}{2}$, depending on the experiment being performed,
where $\vert\theta_1-\theta_2\vert=\frac{\pi}{2}$; a bar over
a phase represents negative phase.
The first and second blocks prepare the
desired basis state $E_i$ and the second block is
implemented only if
$\theta_3=\frac{\pi}{2}$. The shaded block 
represents non-unitary evolution to destroy unwanted
coherences. The third block represents the unitary quantum channel
$\Lambda$ which takes $E_i\rightarrow \Lambda(E_i)$.
The last block achieves the conversion of
the state $\Lambda(E_i)$ 
to determine $\langle E_i \rangle$
by measuring either $ \langle \sigma_{1z} \rangle $ or
$\langle \sigma_{2z} \rangle$. 
(b) NMR pulse sequence for the quantum circuit
given in panel (a). All unfilled rectangles denote $\pi/2$
pulses with the phase written above the pulse, while the solid
rectangle denotes a $\frac{\theta}{2}$ pulse with $\bar{z}$
phase; the evolution time periods are 
$\tau_0=1/(2 J_{\rm CH})$ and $\tau=\theta/(2
\pi J_{\rm CH})$, respectively. The measurement box represents the 
time-domain NMR signal which gives the expectation value of $\sigma_z$
after a Fourier transform. } 
\label{ckt+seq}
\end{figure}
\section{NMR implementation of MSEQPT}
\label{NMRImp}
We demonstrate the MSEQPT protocol on an NMR quantum
information processor for three different unitary processes:
a `no operation' (NOOP), a controlled-NOT (CNOT) gate
and a controlled-Hadamard (CH) gate (we have used the name CH
for this gate where the Hadamard gate is in fact the 
standard pseudo-Hadamard
gate in NMR).  One of the most studied
nonlocal unitary quantum processes is the entangling CNOT
gate, which is a controlled bit flip of the target qubit if
the control qubit is in the state $\vert 1 \rangle$, while
the controlled Hadamard corresponds to applying a Hadamard (or
a pseudo-Hadamard) gate to the target qubit when the
controlled qubit is in the state $\vert 1 \rangle$.
\begin{table}[h!]
\caption{\label{Gates}
Parameters chosen to implement
different unitary quantum processes.}
\begin{ruledtabular}
\begin{tabular}{c c c}
Quantum process &
Phase $\phi $&
$\theta$~~~\\
\colrule
NOOP & $x$, $y$ & 0~~~\\
CNOT & $x$ & $\pi$~~~\\ 
CH & $\overline{y}$ & $\frac{\pi}{2}$~~~\\   
\end{tabular}
\end{ruledtabular}
\end{table}

In Fig.~\ref{ckt+seq} a general rotation through an
angle $\theta$ and a phase $\phi$ on a qubit is represented by
the unitary operator $R_{\phi}^{\theta}$.  Table~\ref{Gates}
lists the values for $\theta$ and $\phi$ used in the quantum
circuit (Fig.~\ref{ckt+seq}(a)) to achieve the desired
unitaries.  NOOP implies `do nothing' or `no operation', the CNOT gate
flips the state of the target qubit (and introduces a phase
$e^{-\iota\frac{\pi}{2}}$) if the control qubit is in the
state $\vert 1\rangle$. The controlled-Hadamard (CH) creates
a superposition state of the target qubit ($\vert 0 \rangle
\rightarrow \vert - \rangle $ and $\vert 1 \rangle
\rightarrow \vert + \rangle $) if the control qubit is in
the state $\vert 1 \rangle$; the states $\vert \pm \rangle =
\frac{1}{\sqrt{2}}(\vert 0 \rangle \pm \vert 1 \rangle)$; a bar
over a phase represents a negative phase.
\begin{figure}[b] 
\includegraphics[angle=0,scale=1.0]{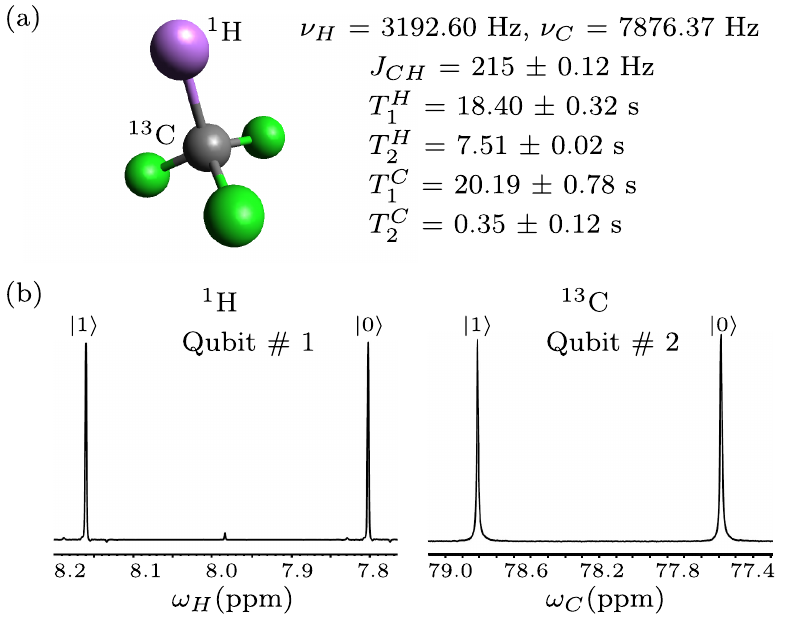}
\caption{(Color online) (a) Molecular structure of
${}^{13}$C labeled chloroform used as a two-qubit quantum
system.  The first and second qubits are encoded as the
nuclear spin ${}^{1}$H and ${}^{13}$C, respectively. The
values of the scalar coupling $J_{\rm CH}$ (in Hz) and
relaxation times $T_{1}$ and $T_{2}$ (in seconds) and
chemical shifts $\nu_i$ are shown alongside.  (b) Thermal
equilibrium NMR spectrum after a $\frac{\pi}{2}$ detection
pulse.}
\label{molecule}
\end{figure}
As discussed in Section~\ref{MSEQPT} we choose the Pauli operators as basis
operators, $\lbrace E_i\rbrace$, although
other choices of basis operators are equally valid as 
the quantum process being tomographed is independent of such
choices. The sixteen product operators for the two-qubit
system are~\citep{ernst-book}: $I$, 
$\sigma_{2x}$,
$\sigma_{2y}$, 
$\sigma_{2z}$, 
$\sigma_{1x}$,
$\sigma_{1x}\sigma_{2x}$, 
$\sigma_{1x}\sigma_{2y}$,
$\sigma_{1x}\sigma_{2z}$, 
$\sigma_{1y}$,
$\sigma_{1y}\sigma_{2x}$, 
$\sigma_{1y}\sigma_{2y}$,
$\sigma_{1y}\sigma_{2z}$, 
$\sigma_{1z}$,
$\sigma_{1z}\sigma_{2x}$, 
$\sigma_{1z}\sigma_{2y}$,
$\sigma_{1z}\sigma_{2z}$, 
where $I$ is a $4 \times 4$ identity matrix,
the $\sigma$ are the Pauli matrices and terms such as
$\sigma_{1x}\otimes\sigma_{2z}$ are written as
$\sigma_{1x}\sigma_{2z}$ for simplicity.  The quantum mapping 
for the experimental measurement
of the expectation values of product operators by 
appropriate single-spin measurements 
is given in Table~\ref{Map}~\citep{singh-pra-16}. For
instance, in order to find the expectation value of $\langle
\sigma_{1x}\sigma_{2y} \rangle $ in the state $\Lambda(E_i)$, we 
map $\Lambda(E_i) \rightarrow \rho_6$ with 
$\rho_6=U_6 \Lambda(E_i) U_6^\dagger$. As per
Table~\ref{Map},
$U_6={\rm CNOT~}\overline{X}_2 Y_1$ which implies that we
need to have the system
undergo a  single-spin  $\frac{\pi}{2}$ rotation
of the first qubit with a phase $y$ and of the second qubit with a phase
$\overline{x}$, followed by a CNOT gate. After this, $\langle
\sigma_{2z} \rangle $ in the state $\rho_6$ is equivalent to
$\langle \sigma_{1x}\sigma_{2y} \rangle $ in the state
$\Lambda(E_i)$.
In the NMR scenario, it is convenient to find 
the expectation values for Pauli $z$-operators as they
correspond to $z$ magnetizations of the nuclear spins.

We encode the two NMR qubits in a molecule of
${}^{13}$C-enriched chloroform dissolved in acetone-D6, with
the nuclear spins ${}^{1}$H and ${}^{13}$C labeled as
`Qubit~1' and `Qubit~2', respectively. The molecular
structure, experimental parameters and the NMR spectrum
obtained at thermal equilibrium after a $\frac{\pi}{2}$
detection pulse are shown in Fig.~\ref{molecule}. 
All the experiments were performed at ambient temperature on
a Bruker Avance III 400 MHz FT-NMR spectrometer equipped
with a BBO probe. 
\begin{figure}[h]
{\includegraphics[scale=1]{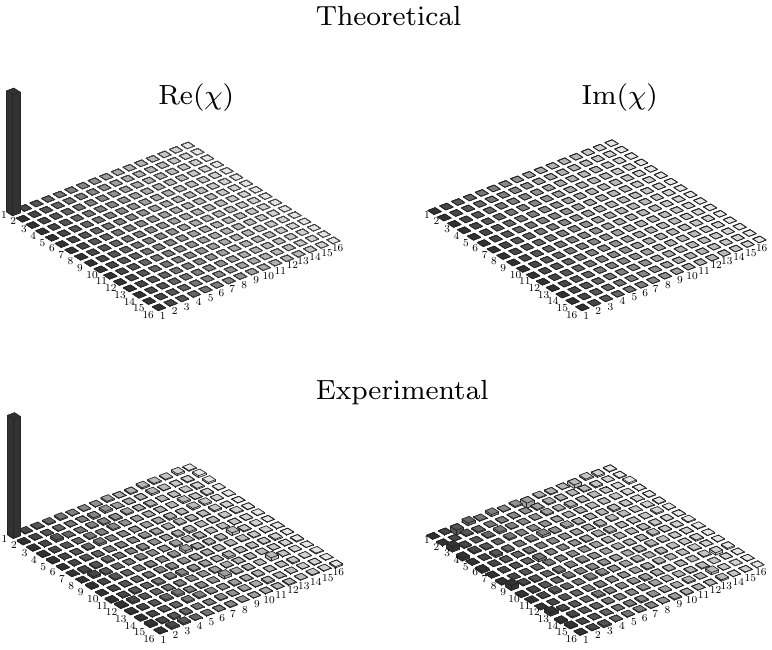}} 
\caption{ The tomographs in the columns
denote the real and imaginary parts of the $\chi$ matrix
respectively, for the NOOP case. The tomographs on
the top represent the theoretically constructed while those
on the bottom represent the experimentally measured $\chi$
matrix for the NOOP case. The fidelity of the
NOOP turned out to be $0.98$.}
\label{tomo-id}
\end{figure}
\begin{figure}[h]
{\includegraphics[scale=1]{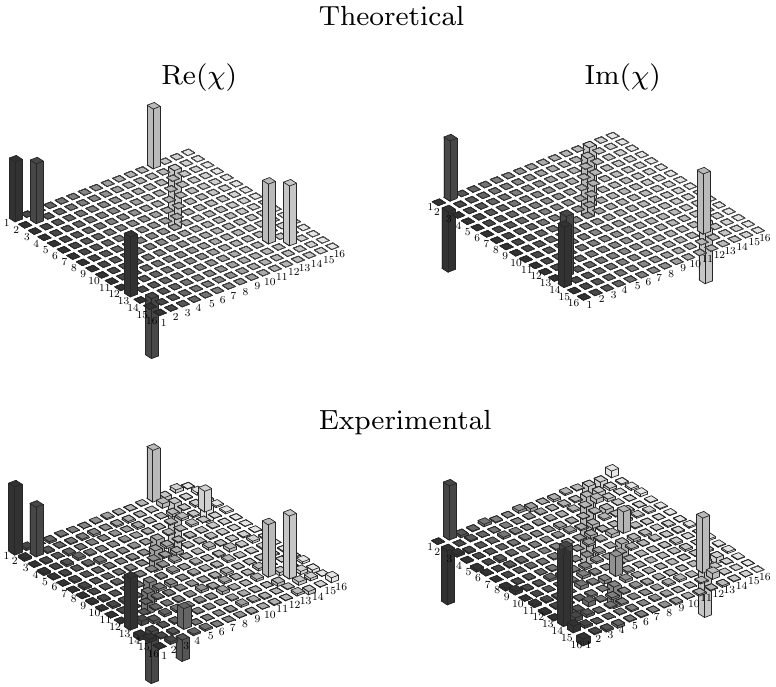}}
\caption{The tomographs in the columns
denote the real and imaginary parts of the $\chi$ matrix
respectively, for the CNOT gate (control-$R_x^{\pi}$).  The
tomographs on the top represent the theoretically
constructed while those on the bottom represent the
experimentally measured $\chi$ matrix of the CNOT operator.
The fidelity of the CNOT operator turned out to be 0.93.}
\label{tomo-cnot} \end{figure}
\begin{figure}[h]
{\includegraphics[scale=1]{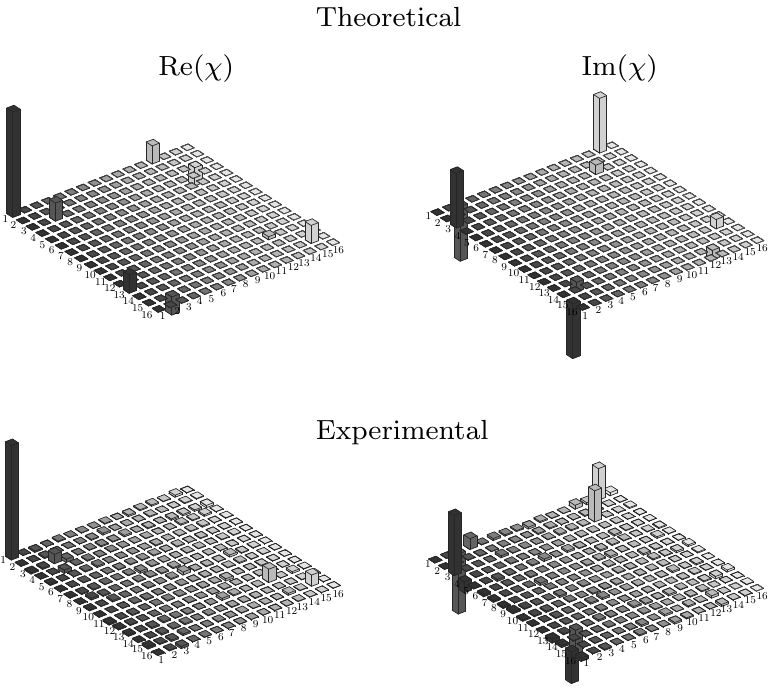}}
\caption{The tomographs in the columns
denote the real and imaginary parts of the $\chi$ matrix
respectively, for the controlled-Hadamard gate
(control-$R_y^{-\frac{\pi}{2}}$). The tomographs on the top
represent the theoretically constructed while those on the
bottom represent the experimentally measured $\chi$ matrix
of the CH operator. The fidelity of the CH operator turned
out to be 0.92.}
\label{tomo-ch}
\end{figure}

The Hamiltonian for a two-qubit system in the rotating frame is given by 
\begin{equation}\label{e_2.13} 
{\mathcal{H}}=2 \pi \left[
(\nu_{{\rm H}}-\nu_{{\rm H}}^{{\rm rf}}) I_z^{{\rm H}} + 
(\nu_{{\rm C}}-\nu_{{\rm C}}^{{\rm rf}}) I_z^{{\rm C}} + 
J_{{\rm CH}} I_z^{{\rm H}} I_z^{{\rm C}}\right]
\end{equation}
where $\nu_{{\rm H}}$, $\nu_{{\rm C}}$ are the 
chemical shifts 
and $I_z^{{\rm H}}$, $I_z^{{\rm C}}$ are the
z-components of the spin angular momentum operators of
the ${}^{1}$H and ${}^{13}$C spins respectively, and J$_{{\rm
CH}}$ is the scalar coupling constant; 
$\nu_{{\rm H}}^{{\rm rf}}$ and
$\nu_{{\rm C}}^{{\rm rf}}$ are the rotating
frame frequencies.
We used the spatial
averaging technique to prepare the spins in an initial
pseudopure state~\citep{cory-physicad,oliveira-book-07}:
\begin{equation} \rho_{00}=\frac{1}{4}(1-\epsilon)I+\epsilon
\vert 00\rangle \langle 00 \vert \end{equation}
where $\epsilon$ is proportional to spin polarization and
can be evaluated from the ratio of magnetic and thermal
energies of an ensemble of magnetic moments $\mu$ in a
magnetic field $B$ at temperature $T$; 
$\epsilon \thicksim
\frac{\mu B}{k_B T}$ and at room temperature  
and for a $B \approx$ 10 Tesla, $\epsilon \approx
\rm{10^{-5}}$.

The quantum circuit and the
corresponding NMR pulse sequence for implementation of the
MSEQPT scheme are shown in Figs.~\ref{ckt+seq}(a) and
\ref{ckt+seq}(b), respectively. 
The circuit is divided into four modules,
separated by dashed blue lines. The
unitary $R_{\phi} ^{\theta}$ in the circuit represents a
local rotation through an angle $\theta$ and phase $\phi$.
All the rotation angles are either zero or $\frac{\pi}{2}$.
The first two modules of the circuit prepare the required basis
state $ E_i $. The shaded rectangle in the first part of the
quantum circuit represents a non-unitary quantum process to
destroy unwanted quantum coherences.
The second module is implemented only if
the experimental settings require a nonzero $\theta_3$. 
The third
model corresponds to the unitary quantum process $\Lambda$ which takes
$E_i \rightarrow \Lambda(E_i)$. The
last modules executes the quantum mapping, as per
Table~\ref{Map}, of the desired operator $ E_i $ to
single-spin 
Pauli $z$-operators. The meter symbol represents an NMR measurement and
only one of the three measurements takes place in one
experimental setting.
The quantum gates
(and local rotations) were implemented using highly accurate
radio frequency (rf) pulses and free evolution periods under
the system Hamiltonian.
Spin selective hard pulses of desired phase were used for
local rotations;
for $\rm{^1H}$ a $\frac{\pi}{2}$ hard
pulse corresponds to an rf pulse of duration 12.95 $\mu$s at 20.19 W
power level while for $\rm{^{13}C}$ the pulse duration was 8.55
$\mu$s at 74.67 W power level.  All the unfilled rectangles
denote $\frac{\pi}{2}$ hard pulses while the filled rectangle is
a $\frac{\theta}{2}$ hard pulse as dictated by the unitary
quantum process $\Lambda $. The phases of all hard pulses are
written above the respective pulse. 
A $z$-gradient was used to
kill the undesired coherences during basis state
preparation. The measurement boxes denote the time-domain NMR
signal which is proportional to the
expectation value of $\sigma_z$ after a Fourier transformation.

The fidelity of experimentally constructed
$\chi_{\rm{expt}}$ with reference to theoretically expected
$\chi_{\rm{theo}}$ was
calculated using the measure~\citep{zhang-prl-14}:
\begin{equation}
{\mathcal F}(\chi^{}_{\rm expt},\chi^{}_{\rm theo})=
\frac{|{\rm Tr}[\chi^{}_{\rm expt}\chi_{\rm theo}^\dagger]|}
{\sqrt{{\rm Tr}[\chi_{\rm expt}^\dagger\chi^{}_{\rm expt}]
{\rm Tr}[\chi_{\rm theo}^\dagger\chi^{}_{\rm theo}]}}
\end{equation} 
Fidelity measure ${\mathcal F}$ is normalized in the sense that as
$\chi_{\rm{expt}} \rightarrow \chi_{\rm{theo}} $ \ie
~experimentally constructed $\chi$ matrix approaches
theoretically expected $\chi$ matrix leads to ${\mathcal
F}$
$\rightarrow$ 1.

The theoretically constructed and experimentally tomographed
$\chi$ matrices for the NOOP, the CNOT
and the CH gates are depicted in
Fig.~\ref{tomo-id}--\ref{tomo-ch}, respectively.  The
fidelity ${\mathcal F}$ for NOOP, CNOT and CH operators was found to
be 0.98, 0.93 and 0.92 respectively.  The upper panel in
Fig.~\ref{tomo-id} depicts the theoretically expected $\chi$
matrix while the lower panel depicts the experimentally
constructed $\chi$ matrix (real and imaginary parts) for the
NOOP case.  Axes of the $\chi$ matrix are labeled by
the indices of the product basis operators $ E_i $.
Similarly Figs.~\ref{tomo-cnot} and \ref{tomo-ch} are the
$\chi$ matrices for CNOT and CH operators respectively. In
all three cases the fidelity F was greater than 0.92 which
signifies the successful experimental implementation of the
MSEQPT protocol. 
\section{Concluding Remarks}
\label{Con}
In this study, we proposed a 
scheme for selective and efficient quantum process
tomography appropriate for NMR systems.  
Our
scheme has a marked advantage in terms of using 
fewer
experiments to determine the selected elements of the
process matrix. We successfully demonstrated the
experimental implementation of the scheme for `no
operation', a controlled-NOT and a controlled-Hadamard gate
on two NMR qubits.  The method is both selective and
efficient and can hence be very useful in any quantum
process which does not require a full experimental
characterization of the
process matrix.  Furthermore, the important task of
calculating the closeness between the implemented process
and the targeted process can be efficiently estimated in a
selective manner. While our modified protocol offers a
clear advantage for the NMR quantum process
tomography experiments, its utility in other
experimental techniques needs to be 
explored further. Efforts are on to implement the
modified protocol for more general quantum
processes and for a larger number of qubits.
\begin{acknowledgments}
All experiments were performed on a Bruker Avance-III 400
MHz FT-NMR spectrometer at the NMR Research Facility at
IISER Mohali. Arvind acknowledges funding from DST
India under Grant No. EMR/2014/000297. K.D. acknowledges
funding from DST India under Grant No. EMR/2015/000556.
\end{acknowledgments}
%
\end{document}